\documentstyle[12pt,a4]{article}
\begin{document}
\begin{titlepage}
\author{E. Dudas $^1$ and J. Mourad $^2$ \\ $^1$ 
Laboratoire de Physique
Th\'eorique et Hautes Energies, \\ B\^at. 211, Univ. Paris-Sud, 
F-91405 Orsay CEDEX, FRANCE \\ $^2$ GPS, Univ. de Cergy-Pontoise,
Site St. Martin, \\ F-95302 Cergy-Pontoise, FRANCE}
\title{\bf On the strongly coupled heterotic string}
\maketitle
\vskip 1cm
\begin{abstract}
We  analyze in detail the  anomaly cancellation conditions for the
strongly coupled $E_8 \times E_8$ heterotic string introduced by Horava 
and Witten and find new features compared to the ten-dimensional 
Green-Schwarz mechanism.
We project onto ten  dimensions the corresponding Lagrangian of the
zero-mode fields.
We find that it has a simple interpretation provided by  the
conjectured  heterotic string/fivebrane duality. The part which
originates from eleven-dimensions is  naturally described in
fivebrane language. 
We discuss physical couplings and scales in four dimensions.

\end{abstract}
\vskip 2cm
\begin{flushright}
LPTHE-Orsay 96/104 \\ GPS 96/110 \\ December 1996 \\ hep-th/9701048
\end{flushright}

\end{titlepage}
\newpage\

\section{Introduction}
It was recently proposed \cite{hw1} that the strong coupling limit of
the $E_8 \times E_8$ heterotic string is described by 
eleven-dimensional supergravity \cite{cjs} compactified on $S_1 / Z_2$,
with the two $E_8$
gauge groups propagating on the two boundaries. This exciting conjecture   
allows to hope for a better understanding of the nonperturbative regime
of the ten and four dimensional heterotic superstring and opens new
perspectives for particle physics string phenomenology.  

In Section $2$, we analyze the analog of the ten-dimensional
 Green-Schwarz anomaly
cancellation mechanism in this context. An important new feature appears,
namely the three-form field $C_{ABC}$, $A,B,C=1 \cdots 10$, which has no
zero modes in ten dimensions generically transforms under gauge
transformations, in
addition to the expected gauge transformation of $C_{11AB}$,
the analog of the antisymmetric tensor field of weakly coupled theory.
 
In  Sections  $3$-$4$, we project onto ten dimensions the Lagrangian 
describing the strongly
coupled $E_8 \times E_8$ heterotic string. The result can be interpreted
in two different ways, by using the membrane origin of the 
eleven-dimensional theory  (membrane quantization condition):

i) in ten-dimensional string units, the resulting Lagrangian is similar
to a mixture of tree-level and string one-loop terms.

ii) in natural $M$-theory units, which we argue to be the ten-dimensional
 fivebrane units, the Lagrangian is similar to a mixture of tree-level
and one-loop fivebrane terms, with all terms originating from 
eleven-dimensions being tree-level.

The natural $M$-theory  units are dual to the string units in the sense
of strong/weak coupling duality. By compactifying from ten to nine dimensions
on a circle we interpret the Kaluza-Klein states and states describing
the wrapping of the membrane around the torus in the fivebrane language.
By compactification from ten to four dimensions, in natural fivebrane units,
we identify in Section $5$ the physical couplings and scales and discuss
the issue of string unification.    
 
Finally some conclusions are drawn together with some comments.
\section{Anomaly cancellation in strongly coupled $E_8 \times E_8$
heterotic string}
The issue of anomaly cancellation was already discussed in \cite{hw1},
\cite{hw2} and \cite{dal}. Due to the new features we encounter, we discuss in
some detail the way it works and compare it to the ten-dimensional 
Green-Schwarz mechanism (GSM).

In the following we use the differential forms notations in eleven dimensions.
For a $p$-form $A$ we define components by the usual definition
$A={1 \over p !} A_{I_1 \cdots I_p} d x^{I_1} \wedge \cdots d x^{I_p}$.
We use the upstairs formulation of \cite{hw2}, e.g. we work on $M_{11}$
directly and impose a $Z_2$ symmetry on the Lagrangian.
In most computations, we use formulae of the type
\begin{equation}
d [\epsilon (x^{11}) A] = 2 \delta (x^{11}) d x^{11} A + \epsilon
(x^{11}) d A \ , \label{eq:ar01}
\end{equation}
where $\epsilon (x^{11})=1$ for $x^{11} >0$ and  $-1$ for $x^{11} <0$
and $A$ is an arbitrary form. It was shown in \cite{hw2}
that the Bianchi identity for the field strength of the three-form $C$
appearing in eleven-dimensional  SUGRA is
(we are only interested in one of the
two ten-dimensional pieces, around, say, $x^{11}=0$) 
\begin{equation}
d G =  a \delta (x^{11}) d x^{11} {\hat I}_4 \ , \label{eq:ar02}
\end{equation}
where we defined $a \equiv {k_{11}^2 / \sqrt 2 \lambda^2}$ as a function
of the eleven-dimensional gravitational coupling $k_{11}$ 
and the gauge coupling
$\lambda$ and ${\hat I}_4 = {1/2} tr  R^2 -tr F^2$. The solution of 
(\ref{eq:ar02}) is to modify the definition of the field strength
\begin{equation}
G = 6 d C - a \delta (x^{11}) d x^{11} Q_3 \ , \label{eq:ar04}
\end{equation}
where $Q_3 = {1/2} \omega_{3L} - \omega_{3Y}$. The value of the
restriction of $G$ around 
the ten-dimensional manifold $M_{10}$ is changed accordingly
\begin{equation}
G |= {a \over 2} \epsilon (x^{11}) {\hat I}_4 + \cdots \ , \label{eq:ar05}
\end{equation}
where $\cdots$ denote terms which vanish for $x^{11}=0$.
Compatibility of (\ref{eq:ar04}) and (\ref{eq:ar05}) gives the
value of the restriction of the three-form on the ten-dimensional boundary:
\begin{equation}
C| = {a \over 12} \epsilon (x^{11}) Q_3 + d \Omega \ , \label{eq:ar050}
\end{equation} 
where $\Omega$ is a two-form field which can be written, near $M_{10}$,
by $Z_2$ symmetry  as
$\Omega = \epsilon  (x^{11}) \Omega'$, with $\Omega'$ a two-form living
on $M_{10}$.
The term responsible for the anomaly cancellation is the
Chern-Simons term \cite{cjs}
\begin{equation}
W = -{ \sqrt 2 \over k_{11}^2} \int_{M_{11}} C \wedge G \wedge G
\ . \label{eq:ar06}
\end{equation}
The variation $\delta C$ is found by imposing $\delta G =0$ in 
(\ref{eq:ar04}). By using $\delta Q_3 = d Q_2^1$, where
$Q_2^1$ is a two-form linear in the gauge transformation parameter, we
find that the result can be parametrized by a free parameter $\alpha$
and reads
\begin{equation}
\delta C = \alpha \delta C_1 + (1-\alpha ) \delta C_0 + d
\Lambda_{\alpha} \ , \label{eq:ar07}
\end{equation}
where $\Lambda_{\alpha}$ is a two-form and we defined
\begin{equation}
\delta C_1 \equiv {a \over 12} \epsilon (x^{11}) d Q_2^1 \ , \ \ 
\delta C_0 \equiv -{a \over 6} \delta (x^{11}) d x^{11} Q_2^1 \ . 
\label{eq:ar08}
\end{equation}
Identifying the gauge transformation of (\ref{eq:ar050}) with the
restriction of (\ref{eq:ar07}) to $M_{10}$ we find that
\begin{equation}
\delta \Omega'| = -{a \over 12} (1-\alpha) Q_2^1 + \Lambda_{\alpha}
\ . \label{eq:ar081}
\end{equation} 
Notice that $W$ is invariant under the usual three-form gauge
transformations $\delta C = d \Lambda_{\alpha}$ only
if $\Lambda_{\alpha}=0$ on $M_{10}$, which will be assumed in the
following.
It is interesting that, even by putting $\Lambda_{\alpha}=0$, $\delta C$
depends on $\alpha$  as a total derivative
\begin{equation}
\delta C = \delta C_0 + \alpha {a \over 12} d [\epsilon (x^{11}) Q_2^1 ]
= \delta C_1 - {a \over 12} (1-\alpha) d [\epsilon (x^{11}) Q_2^1 ] 
\ . \label{eq:ar080}
\end{equation}
However, the two-form in the bracket does not vanish on $M_{10}$ and 
therefore, as (\ref{eq:ar06}) is not invariant under the $3$-form gauge
transformations, the value of the anomaly depends on $\alpha$.
The gauge variation of $W$ can be splitted into two pieces:
\begin{equation}
\delta W = \alpha \delta_1 W + (1-\alpha ) \delta_0 W \ .
\label{eq:ar09}
\end{equation}
By using (\ref{eq:ar02}), (\ref{eq:ar05}) and (\ref{eq:ar08}) we find
immediately
\begin{equation}
\delta_0 W = {\sqrt 2 a^3 \over 24 k_{11}^2 } \int_{M_{10}} Q_2^1 
\wedge {\hat I}_4  \wedge {\hat I}_4  \ . \label{eq:ar010}
\end{equation}
In order to find $\delta_1 W$ we insert (\ref{eq:ar08}) into
(\ref{eq:ar06}),
integrate by
parts,  use the formulae (\ref{eq:ar01}), (\ref{eq:ar02}) and (\ref{eq:ar05}). 
The result is
\begin{equation}
\delta_1 W = {\sqrt 2 a^3 \over 8 k_{11}^2 } \int_{M_{10}} 
Q_2^1  \wedge {\hat I}_4  \wedge {\hat I}_4  \ . \label{eq:ar011}
\end{equation} 
On the other hand, the ten-dimensional $E_8 \times E_8$  anomaly 
is  (see, for example, \cite{gsw}) 
\begin{equation}
\delta \Gamma = {1 \over 48 (2\pi )^5 } \int_{M_{10}} 
Q_2^1 \wedge [{1 \over 4}{\hat I}_4\wedge {\hat I}_4 -X_8 ]   \ , \label{eq:ar012}
\end{equation}   
where $X_8 \equiv -{1 \over 8} tr R^4 + {1 \over 32} (tr R^2)^2$.
By comparison of (\ref{eq:ar09}) and (\ref{eq:ar012}) we find that the
${\hat I}_4\wedge {\hat I}_4$ part of the
anomaly is cancelled provided the following relation holds
\begin{equation}
{k_{11}^4 \over \lambda^6 } =-{1 \over 4 (1+2 \alpha ) (2 \pi )^5 } \ .
\label{eq:ar013}
\end{equation}
The remaining anomaly cancellation can be achieved from another term in
the $M$ theory Lagrangian \cite{da}, which can be viewed as a term
cancelling fivebrane worldvolume anomalies \cite{dlm} or, by
compactifying one dimension, as a one-loop term in the type $IIA$ superstring 
\cite{vw}. In our conventions, it reads
\begin{equation}
W_5 = -{T_3 \over 2 \sqrt 2 (2 \pi )^4 } \int_{M_{11}} C \wedge X_8 \ ,
\label{eq:ar014}
\end{equation}
where $T_3$ is the membrane tension related to ${\alpha}'$ by
$T_2=2 \pi {\alpha}'^{1/2} T_3 = {1 \over 2 \pi {\alpha}'}$.
We  use the membrane quantization condition \cite{dlm}, \cite{da}, \cite{witt}
\begin{equation}
{{(2 \pi)}^2  \over k_{11}^2 T_3^3} = 2  m \ , \ m = integer \ or \
halfinteger \ . \ 
\label{eq:ar015}
\end{equation}  
The gauge and Lorentz variation of $W_5$ is then easily computed to be
\begin{equation}
\delta W_5 = {a \sqrt 2 \over 24 (2 \pi )^{10/3} (2m k_{11}^2)^{1/3}} 
\int_{M_{10}} Q_2^1 \wedge X_8 \ .
\label{eq:ar017}
\end{equation}
Notice that, due to (\ref{eq:ar080}) and $d tr R^2 =d tr R^4 = 0$,
$\delta W_5$  is independent of $\alpha$.
The result is that $\delta W + \delta W_5 + \delta \Gamma=0$ if the
gauge and the eleven-dimensional gravitational couplings are related by the
relation
\begin{equation}
m^{1/3} \lambda^2 = 2  \pi (4 \pi k_{11}^2 )^{2/3} \ . \label{eq:ar019}
\end{equation} 
The final step is to compare (\ref{eq:ar013}) with (\ref{eq:ar019}).
We find that the gauge anomaly is completely cancelled if the membrane
quantization integer (or half-integer) parameter $m$ is related to $\alpha$ by
\begin{equation}
\alpha = -{1 \over 2} (1 + {1 \over m }) \ . \label{eq:ar020}
\end{equation}
For example, $m=1$ implies $\alpha = -1$.
The relation (\ref{eq:ar019}) has a physical meaning only for $m >0$. 
A puzzling result is that the value $\alpha =0$, which is the most
natural value from a weak coupling ten-dimensional string point of view is
only obtained for the unphysical value $m=-1$. To be more precise, we
rewrite the gauge transformations (\ref{eq:ar07}) in component fields 
(with $Q_2^1 = -tr \epsilon F$, $\epsilon$ being the gauge transformation
parameter):
\begin{eqnarray}
\delta C_{ABC} &=& -\alpha {a \over 12} \epsilon (x^{11}) \left[
\partial_A (tr \epsilon F_{BC}) \pm 2 \ perm. \right] \ , \\
\delta C_{11AB}&=& (1-\alpha ) {a \over 6} \delta (x^{11}) tr \epsilon 
F_{AB} \ , \label{eq:ar021}
\end{eqnarray}
where $A,B,C$ are ten-dimensional indices. The contribution of $\delta
C_{ABC}$ to the gauge anomaly has no weak coupling heterotic string
interpretation and would disappear only for $\alpha =0$. Precisely in this case
we could identify $C_{11AB}$ with the ten-dimensional string antisymmetric
tensor field $B_{AB}$. However, this corresponds, by using
(\ref{eq:ar019}) to an imaginary gauge coupling.

A more general framework for comparing with the weakly coupled GSM consists
of using a more  general modification of the four-form
field strength compatible with the Bianchi identity (\ref{eq:ar02})
\begin{equation}
G=6 d C -a \beta \delta (x^{11}) d x^{11} Q_3 + {a \over 2} (1 -\beta)
\epsilon (x^{11}) {\hat I}_4 \ , \label{eq:ar022}
\end{equation}   
where $\beta$ is an arbitrary real parameter. In this case, the
eqs. (\ref{eq:ar019}) and (\ref{eq:ar020}) become
\begin{equation}
m^{1/3} \lambda^2 = 2  \pi \beta (4 \pi k_{11}^2 )^{2/3}, \ \ 
\alpha = -{1 \over 2} (1 + {\beta^2 \over m }) \ . \label{eq:ar023}
\end{equation} 
The case corresponding to the weakly coupled GSM corresponds to the
set of values $\alpha=0, \beta =-1, m=-1$.  This more general approach
induces additional corrections to the SUSY transformation law for $C$:
\begin{equation}
{\tilde \delta} C = -{a \over 12} \beta \alpha \epsilon (x^{11}) d (tr
A {\tilde \delta} A) + {a \over 6} \beta (1-\alpha) \delta (x^{11}) d x^{11} tr
A {\tilde \delta} A \ , \label{eq:ar024}
\end{equation}
which imply an additional correction to the SUSY transformation
law for $G$ compared to \cite{hw2}. It reads
\begin{equation}
{\tilde \delta} G = a \left[ 2 \beta \delta (x^{11}) d x^{11} tr
(F {\tilde \delta} A) - (1- \beta) \epsilon (x^{11}) tr
(F {\tilde \delta} F) \right]   \ , \label{eq:ar0241}
\end{equation}
and this asks for additional terms in the 
fermionic part of the Lagrangian in \cite{hw2} in order to preserve
supersymmetry. This is a  viable alternative, but we will not pursue
further this observation here.  
Therefore, it seems that even for $\beta =1$ the anomaly cancellation is 
essentially different in strong
coupling regime of the heterotic string, which is a puzzle in our
usual understanding of anomalies.

Our conclusions for the minimal
choice $\beta=1$ in (\ref{eq:ar023}) differ from that in \cite{hw2}.
We find that $C_{ABC}, A,B,C = 1 \cdots 10$ must transform under
gauge transformations for a real gauge coupling. Correspondingly, the GSM 
does not directly reduce to the ten-dimensional case. 
Comparing with the downstair 
(manifold with  boundary) approach, it is interesting to note that the gauge
transformations (\ref{eq:ar07}), imposed by the modification of the 
corresponding field strength, are different from those of
\cite{dal}. The downstair gauge 
transformations of \cite{dal} are obtained for $\Omega=0$ in
(\ref{eq:ar050}), but this choice is incompatible with (\ref{eq:ar081}).
This conclusion holds in the  more general case described in (\ref{eq:ar022}).
\footnote{We can recover our results in the downstair approach by using
$C|={a \over 12} Q_3 + d \Omega'$, (\ref{eq:ar081}),(\ref{eq:ar020}) and using
the integration by part formula $\int_{M_{11}} d \omega =
\int_{x^{11}= \pi} \omega - \int_{x^{11}=0} \omega$.}    
\section{Strongly coupled $E_8 \times E_8$ heterotic string Lagrangian
viewed from ten dimensions}

We try now to interpret different terms of the heterotic Lagrangian in the 
strongly coupled regime in a heterotic weak coupling language, by using
the membrane quantization condition (\ref{eq:ar015}), which allows us
to relate the string slope ${\alpha}'$ to $k_{11}$ by \cite{da}  
\begin{equation}
{\alpha}' = {  \left[ {2 m k_{11}^2 \over {(2
\pi)}^8}   \right] }^{2 \over 9} \ . \label{eq:ar60}
\end{equation}
We certainly do not want to propose to reduce the $M$-theory physics to a
ten-dimensional physics, just to show that projecting the Horava-Witten 
Lagrangian
onto ten-dimensions we can naturallly interpret the result in terms of 
heterotic weak coupling variables.
In the following we only consider zero modes of the fields in a 
Kaluza-Klein decomposition, which are interpreted as the usual ten-dimensional
heterotic fields in the weak coupling description. For this, we develop
eleven-dimensional fields in a Fourier expansion, for example
\begin{equation}
C_{11AB} (x^{11},x) = \sum_{n=0}^{\infty} cos{n x^{11} \over {\alpha'}^{1/2}}
C_{11AB}^{(n)} (x) \ , \label{eq:ar61}
\end{equation}
where $x=x_1 \cdots x_{10}$ and keep into account only $C^{(0)}$.
Similarly, we expand
\begin{equation}
\delta (x^{11}) = {1 \over 2 \pi {\alpha'}^{1/2}} + 
{1 \over \pi {\alpha'}^{1/2}} \sum_{n=1}^{\infty} 
cos{n x^{11} \over {\alpha'}^{1/2}} \ . \label{eq:ar62}
\end{equation}
Then, by using (\ref{eq:ar019}), (\ref{eq:ar60}), (\ref{eq:ar61}), 
(\ref{eq:ar62}) we can write the zero-mode part of eq. (\ref{eq:ar04}) as
(we set $\omega_{3L}=0$ in the following, as all our considerations below
can be trivially generalized by the inclusion of higher-order
gravitational terms) 
\begin{equation}
G^{(0)} = 6 d C^{(0)} + {{\alpha'} \over 2 \sqrt 2} d x^{11} \omega_{3Y} 
\ , \label{eq:ar63}
\end{equation}
which is similar to the ten-dimensional relation $H=d B -{{\alpha'} \over 2}
\omega_{3Y}$.  Notice that the membrane quantization parameter $m$
cancelled out in the final expression. 

The bosonic part of the strongly coupled $E_8 \times E_8$ heterotic
string Lagrangian is \cite{hw2}
\begin{eqnarray}
&& L={1\over \kappa_{11}^2}\int_{M^{11}}d^{11}x\sqrt g
\left(-{1\over 2}R^{(11)} 
-{1\over 48}G_{IJKL}G^{IJKL}  \right) \label{eq:ar1} \\
&& -{\sqrt 2 \over 3456 \kappa_{11}^2} \int_{M^{11}} d^{11}x
\epsilon^{I_1\dots I_{11}}C_{I_1I_2I_3}G_{I_4\dots I_7}G_{I_8\dots
I_{11}}- {1 \over \lambda^2} \int_{M^{10}}d^{10}x
\sqrt g \frac{1}{4} tr F_{AB} F^{AB} \ , \nonumber 
\end{eqnarray}
where $I,J,K,L = 1 \dots 11$.

We will be mainly interested in the projection of this Lagrangian onto one
of its two boundaries, let's say $x^{11}=0$. The  physical
ten-dimensional bosonic
fields, which are invariant under the $Z_2$ orbifold projection are 
$g_{11,11},g_{AB},A_B$ and $C_{11AD}$.
Moreover, by
considerations explained in \cite{w}, in the ten-dimensional string metric
$g_{st,AB}=e^{2 \phi \over 3 }g_{AB}$ we have $g_{11,11}=e^{4 \phi /3}$ ,
where $\phi$ is the heterotic string dilaton. 
In the following, all the fields with an eleven-dimensional 
origin mean implicitly
the corresponding zero-modes part.
For the zero mode of $G_{ABCD}$ we use (\ref{eq:ar05}).
We neglect in the  following the Green-Schwarz gravitational anomaly
terms which are irrelevant for our purposes.
 
The result in the string metric is
\begin{eqnarray}
&& L^{(10)} = {2\pi {\alpha}'^{1/2} \over
\kappa_{11}^2}\int_{M^{10}}d^{10}x \sqrt g_{st} e^{-2 \phi}
\left[ -{1\over 2} R^{(10)} -4 {(\partial \phi)}^2
-{1\over 12} G_{11ABC}^2  \right. \nonumber \\
&&  \left. -{3 \kappa_{11}^{4/3} \over 8 
{(4 \pi)}^{10/3}}
e^{2 \phi} {(F^2)}^2 
- {\sqrt2 \kappa_{11}^{4/3} m^{2/3} \over 64 {(4 \pi)}^{10/3} \sqrt
g_{st}} e^{2 \phi}
C_{11} F^2 F^2 \right] \label{eq:ar5} \\
&& - {m^{1/3} \over 2\pi(4\pi\kappa_{11}^2)^{2/3}} \int_{M^{10}}d^{10}x
\sqrt g_{st}\,\left( \frac{1}{4} e^{-2 \phi} tr F_{AB} F^{AB} \right), 
\nonumber
\end{eqnarray}
where we defined ${(F^2)}^2 \equiv F_{[AB}^a F_{CD]}^a F^{b,[AB} F^{b,CD]}$ and

\noindent $C_{11} F^2 F^2 \equiv \epsilon^{A_1 \dots A_{10}}C_{11A_1A_2} 
F_{[A_3A_4}^a F_{A_5A_6]}^a F_{[A_7A_8}^b F_{A_9A_{10}]}^b$.
The Lagrangian $L^{(10)}$ can be written in a more transparent way by
using heterotic string  notations. 
We define the ten-dimensional Newton constant by 
$k_{11}^2 = 2 \pi {\alpha}'^{1/2} k_{10}^2$ 
and use (\ref{eq:ar60}). Then $L^{(10)}$ reads
\begin{eqnarray}
&& L^{(10)}={1 \over 2
\kappa_{10}^2}\int_{M^{10}}d^{10}x \sqrt g_{st} e^{-2 \phi}
\left[-R^{(10)} -4 {(\partial \phi)}^2
-{1\over 6} G_{11ABC}^2 -{{\alpha}' \over 4} tr F_{AB} F^{AB}  \right. 
\nonumber \\
&& \left. -{3 m k_{10}^2 \over 32 {(2 \pi)}^5 {\alpha}'}
e^{2 \phi} {(F^2)}^2 \right] - {m \over 2569 \sqrt 2 {(2 \pi)}^5
{\alpha}'} \int_{M^{10}}d^{10}x C_{11} F^2 F^2 \ . \label{eq:ar7}
\end{eqnarray} 
This expression is very similar to a weakly-coupled
heterotic string Lagrangian
containing a mixture of tree-level and one-loop terms, where the
one-loop terms in (\ref{eq:ar7}) have an additional factor of 
$e^{2 \phi}$ compared to the tree-level terms. 
By the identification $6 \sqrt 2 C_{11AB}^{(0)}=B_{AB}$,
$ \sqrt 2 G_{11ABC}^{(0)}=H_{ABC}$, needed in order to identify (\ref{eq:ar63})
with the string relation  $H=d B -{{\alpha'} \over 2}
\omega_{3Y}$, we get the usual
Green-Schwarz term for $m=-1$, in agreement with the conclusions of the
preceeding paragraph.

The analogy with a weakly-coupled heterotic Lagrangian becomes stronger
by looking in more detail at the one-loop 
terms of the type $F^4$ in the ten-dimensional $E_8 \times E_8$
heterotic string. They were computed long-time ago in \cite{ejm} and
were found to be proportional to
${1 \over {(2 \pi)}^5 {\alpha}'} t^{ABCDEFGH} trF_{AB} F_{CD} F_{EF}
F_{GH}$, where the tensor
$t^{ABCDEFGH}$ is defined by the expression 
\begin{eqnarray}
&&{t^{ABCDEFGH}}=-{1\over 2} (g^{AC} g^{BD} - g^{AD}
g^{BC}) (g^{EG} g^{FH} - g^{EH} g^{FG}) \nonumber \\
&&-{1\over 2} (g^{CE} g^{DF} - g^{CF} g^{DE}) (g^{GA} g^{HB} -
 g^{GB} g^{HA}) -{1\over 2} (g^{AE} g^{BF} -g^{AF} g^{BE}) (g^{CG} g^{DH} -
 g^{CH} g^{DG}) \nonumber \\
&&+{1\over 2} (g^{BC} g^{DE} g^{GF} g^{HA} + g^{BE} g^{FC} g^{DG} g^{HA}
+ g^{BE} g^{FG} g^{CH} g^{DA} + \,{\rm perm.}). \label{eq:ar8}
\end{eqnarray}
By a rather straightforward algebra one can prove
the following formula
\begin{eqnarray} 
&&t^{ABCDEFGH} Tr F_{AB} F_{CD} F_{EF} F_{GH} =  
{1 \over 100} t^{ABCDEFGH} (Tr F_{AB} F_{CD}) (Tr F_{EF} F_{GH})
\nonumber \\
&&= {1 \over 25} \left[ - (Tr F^{AB} F^{CD})(Tr F_{AB} F_{CD})+
2 (Tr F^{AB} F^{CD})(Tr F_{BC} F_{DA}) \right] \nonumber \\
&&+ {1 \over 50} \left[ (-Tr F^{AB} F_{AB})(Tr F^{CD} F_{CD}) +
2 (Tr F^{AB} F_{BC})(Tr F^{CD} F_{DA}) \right] \ , \label{eq:ar9}
\end{eqnarray}
where $Tr$ is the trace in the adjoint representation (which is
also the fundamental representation) of $E_8$.
On the other hand, the $F^4$ term appearing in (\ref{eq:ar7}) can be
rewritten as
\begin{equation}
9 {(F^2)}^2 = 
3 (Tr F_{AB} F_{CD}) (Tr F^{AB} F^{CD})  + 6 (Tr F_{AB} F_{CD})
Tr F^{AD} F^{BC} \ . \label{eq:ar10}
\end{equation}
Comparing (\ref{eq:ar9}) and (\ref{eq:ar10}) we conclude that the $F^4$
term  appearing in the strongly coupled Horava-Witten heterotic
Lagrangian projected onto ten-dimensions is indeed similar to a
one-loop effect in the weakly-coupled heterotic string, but the
numerical factor in (\ref{eq:ar7}) is (for $m=1$) three times bigger.
Still, it is interesting that the combination in eq. (\ref{eq:ar10})
already appears in (\ref{eq:ar9}).
Actually, a certainly  better way to define a ten-dimensional
Lagrangian is to
integrate over the Kaluza-Klein modes. This could perhaps
reproduce correctly the one-loop result\footnote{This would be
important for the proposal made in \cite{bfss}. We thank S.P. de Alwis
for this remark}, but this is
beyond the goal of this letter (see for ex. \cite{ckm} for an
integration from five to four dimensions in the same context).
Consequently, from a weak-coupling
point of view, the projected Lagrangian is similar to a mixture of
tree-level and one-loop terms, but with value of the couplings
renormalized compared to their weak-coupling values. 
Notice, however,
that all this picture is rather formal because we are in a strongly
coupled regime for the heterotic string where a perturbative Lagrangian
in the string coupling does not have too much sense.
In particular, we believe  that it is not justified to further
compactify to four-dimensions in string units in order to define
four-dimensional couplings and scales.  
A better picture is obtained by going to the dual
fivebrane picture to be analyzed in the next paragraph.  
\section{Fivebrane picture}
It is well known  that the  eleven-dimensional supergravity can be
obtained from the worldvolume action of the eleven-dimensional
supermembrane by imposing the Kappa symmetry \cite{bst}, giving a
possible membrane  origin of M-theory. On the other hand, there are
arguments by Duff et al. \cite{dlm} in favor of a 
membrane/fivebrane duality in
eleven dimensions. In this section we argue that the part of 
$L^{(10)}$ originating from eleven-dimensions is naturally described in
ten-dimensional fivebrane units by using arguments very similar to that
used by Duff and coll. \cite{duff} in order to support the heterotic /
fivebrane duality in ten dimensions. Indeed, our form for $L^{(10)}$ is
very  close to the low-energy fivebrane  Lagrangian of Duff and coll., 
except that the $F^4$ terms appearing in (\ref{eq:ar7})
are only a part of the terms  conjectured in \cite{duff}. 
We therefore follow closely their analysis trivially adapted to our case .
The fivebrane tension ${\beta}'$ is related to the string tension by the
relation
\begin{equation}
m {(2 \pi)}^5 {\alpha}' {\beta}' = 2 k_{10}^2 \ . \label{eq:ar11}
\end{equation}
By using the membrane quantization condition (\ref{eq:ar015}) we get the
analog of (\ref{eq:ar60}) for the fivebrane tension
\begin{equation}
{\beta}' = { \left [{2 k_{11}^2 \over {(2 \pi)}^{5} m^2} \right] }^{2 \over
3} \ . \label{eq:ar12}
\end{equation}
By performing a Weyl transformation $g_{st}=e^{2 \phi \over 3 }g_5$,
where $g_5$  is the ten-dimensional fivebrane metric and performing
the Hodge duality $K= \sqrt 2 e^{-\phi} { ^\ast G_{11}}$, with $(G_{11})_{ABC}
\equiv G_{11ABC}$, we get the Lagrangian
\begin{eqnarray}
&& L^{(10)}={1 \over 2
\kappa_{10}^2}\int_{M^{10}}d^{10}x \sqrt g_{5} e^{2 \phi \over 3}
\left[-R^{(10)}
-{1\over 12} K^2 - {3 m^2 {\beta}' \over 64 } {(F^2)}^2 \right. 
\nonumber \\  
&& \left.  - {k_{10}^2 \over 2 {(2 \pi)}^5 m {\beta}'} e^{-{2 \phi
\over 3}} tr F_{AB} F^{AB} \right] \ . \label{eq:ar13}
\end{eqnarray}
The tree-level field equations for the three-form, written in form language
\begin{equation}
d ^\ast G = - {1 \over \sqrt 2} G \wedge G \ , \label{eq:ar70}
\end{equation}
after projection onto  ten-dimensions and use of
 (\ref{eq:ar05}),(\ref{eq:ar019}) and
(\ref{eq:ar12}) read, in fivebrane units
\begin{equation}
d K = - m^2  \beta' tr F^4 \label{eq:ar71}
\end{equation}
and become the Bianchi identity for $K$.
Notice that all the
terms  with an eleven-dimensional origin in (\ref{eq:ar7}) are
tree-level if the fivebrane
coupling is $\lambda_5 = e^{-{\phi \over 3}}$.  A notable exception is the
usual Yang-Mills term $F^2$ (of a ten dimensional origin), which is now 
one-loop (whatever that means for a five-brane). Also, $g_5$ is actually the
natural  metric for the M-theory. More precisely, the fivebrane
units in which the dilaton has no kinetic term are obtained from 
the eleven dimensional supergravity in the natural units 
$g_{11,11}=e^{4 \phi \over 3},g_{AB}^{(11)}=g_{AB}^{(10)}$.

An additional argument for heterotic/fivebrane duality can be obtained
by compactifying one additional coordinate, $x_{10}$. As argued in \cite{hw1},
in the M-theory metric the masses of states in the compactified
theory are given by
\begin{equation}
M^2 = {l^2 \over R_{11}^2} + {m^2 \over R_{10}^2} + n^2 R_{10}^2
R_{11}^2 \ , \label{eq:ar14}
\end{equation}
where $l,m$ are Kaluza-Klein modes and $n$ describes the wrapping of the
membrane around the two-torus. These states should have an
interpretation in the heterotic and in the fivebrane picture. 
In the string metric, they become
\begin{equation}
M_h^2 = {l^2 \over {\lambda_{E_8}^2}} + {m^2 \over R_{E_8}^2} + n^2 R_{E_8}^2
 \ , \label{eq:ar15}
\end{equation}  
where $R_{11} = {\lambda_{E_8}}^{2/3}$ and $R_{10}= {R_{E_8} \over 
{\lambda_{E_8}}^{1/3}}$. In the fivebrane metric, we get analogously
\begin{equation}
M_5^2 = {l^2 {\lambda_{5}^4}} + {m^2 \over R_{5}^2} + n^2 {R_5^2 \over
{\lambda_{5}^4}}
 \ , \label{eq:ar16}
\end{equation}
where $R_5=R_{10}= {R_{E_8} / {\lambda_{E_8}}^{1/3}}$ and
${\lambda_{5}}={\lambda_{E_8}}^{-1/3}$. The natural interpretation of
these expressions is that nonperturbative string states labeled by $l$
become perturbative in fivebrane picture. The string winding states $n$
become windings of the string solitonic solution around the compactified
torus and Kaluza-Klein $m$ states keep their
interpretation in the dual picture.

The Lagrangian (\ref{eq:ar13}) is
in the perturbative regime when the heterotic string is strongly coupled
and can eventually be compactified to four dimensions. 

\section{Couplings and scales in four dimensions}
We  now compactify to four dimensions in order to define physical
quantities relevant for phenomenology.\footnote{The following
considerations apply in the lowest order in $k_{11}^{2/3}$, in which
case we can approximate the compactified space by $K \times S_1 / Z_2$,
where $K$ is a Calabi-Yau manifold.} 
A special attention will be paid
for writing four-dimensional relations in the 
appropiate metric for the weak and
strong coupling  respectively and for using four-dimensional 
fields $S$ and $T$
defined in the Witten truncation \cite{wit}. 
We follow the  conventions and notations of \cite{wi}. For the
weakly-coupled heterotic string, the ten-dimensional
terms we are interested in then read
\begin{equation}
L_h^{10} = - \int d^{10}x \sqrt g_{st} e^{-2 \phi} \left[ {4 \over
{\alpha '}^{4}}R + {1 \over {\alpha '}^{3}} tr F^2 + \cdots \right]
\ . \label{eq:ar17}
\end{equation} 
In the following, six-dimensional  compact indices will be denoted by latin
letters $i,j$ and four-dimensional indices by greek letters $\mu , \nu$. 
Four dimensional string units are obtained by setting
\begin{equation}
g_{ij}^{10}=e^{\sigma} \delta{ij} \ , \ g_{\mu \nu}^{10}= g_{\mu \nu}
\ . \label{eq:ar18}
\end{equation}
By defining the grand unification scale $M_{GUT}$ by
$V_6^{-1}=M_{GUT}^6$, where $V_6$ is the volume of the compactified
space, we get for the Newton constant $G_N$ and GUT coupling
$\alpha_{GUT}$ in four dimensions :
\begin{equation}
16 \pi G_N = {t^3 \over 4 s} {\alpha '}^{4} M_{GUT}^6 \ , \
16 \pi {\alpha}_{GUT} = {t^3 \over s} {\alpha '}^{3} M_{GUT}^6 \ , 
\label{eq:ar19}
\end{equation}
where $s = e^{-2 \phi} e^{3 \sigma}, t=e^{\sigma}$. By  combining the two 
equations, we get the well known relation
\begin{equation}
G_N = {1 \over 4} {\alpha}' {\alpha}_{GUT} \ . \label{eq:ar20}
\end{equation}
In the weak-coupling regime $e^{2 \phi}< 1$ which, by using (\ref{eq:ar19})
and (\ref{eq:ar20}) reads
\begin{equation}
G_N > {({ \pi \over 4})}^{1/3} {{\alpha}_{GUT}^{4/3} \over M_{GUT}^2}
\ , \label{eq:ar21}
\end{equation}
which is too large and gives rise to the so-called string unification
problem \cite{vadim}, \cite{string}. 

The situation is different for the strongly-coupled regime. It is true
that, viewed from ten dimensions the Lagrangian (\ref{eq:ar7}) is formally 
similar to a
perturbative string one. However, due to the strong coupling it
is more natural to work in the dual fivebrane coordinates. For notational
convenience, we still use ${\alpha}'$ instead of ${\beta}'$ in our
expressions. The relevant terms in the ten-dimensional Lagrangian are
\begin{equation}
L_5^{10} = - \int d^{10}x \sqrt g_{5} e^{2 \phi \over 3} \left[ {4 \over
{\alpha '}^{4}}R + e^{-{2 \phi \over 3}} {1 \over {\alpha '}^{3}} tr F^2 + \cdots \right]
\ . \label{eq:ar22}
\end{equation}
The four-dimensional fivebrane units are obtained by setting \cite{pierre}
\begin{equation}
g_{5,ij}^{10}=e^{\sigma} \delta{ij} \ , \ g_{5, \mu \nu}^{10}= e^{-2
\sigma} g_{\mu \nu}, \label{eq:ar23}
\end{equation}
and the four-dimensional $Re \ S$ and $Re \ T$ fields are defined by
\begin{equation}
s = e^{3 \sigma} \ , \ t = e^{\sigma} e^{2 \phi \over 3} \ . \label{eq:ar24}
\end{equation}
In this case, by compactifying (\ref{eq:ar22}) to four dimensions we get
\begin{equation}
16 \pi G_N = {s \over 4 t} {\alpha '}^{4} M_{GUT}^6 \ , \
16 \pi {\alpha}_{GUT} = {\alpha '}^{3} M_{GUT}^6 \ . \ 
\label{eq:ar25}
\end{equation}
By combining them, we get the  analog of (\ref{eq:ar20}) and
(\ref{eq:ar21})
\begin{equation}
G_N = {s \over 4 t} {\alpha}' {\alpha}_{GUT} = 
{({ \pi \over 4})}^{1/3} {s \over t} {{\alpha}_{GUT}^{4/3}
\over M_{GUT}^2} \ . \label{eq:ar26} 
\end{equation} 
The four-dimensional 
 fivebrane coupling is $t^{-1/2}$ which means $t$ has large
values in fivebrane weak coupling / string strong coupling regime.
Clearly this is exactly what we need in (\ref{eq:ar26}) in order to
acommodate the small observed value of the Newton constant $G_N$.
Notice that the threshold associated with the eleventh dimension
\begin{equation}
M_{11} \equiv {1 \over R_{11}} = {s^{1/3} \over t} << M_{GUT}
= e^{-\sigma \over 2} = {1 \over s^{1/6}} \  \label{eq:ar 27}
\end{equation}
is indeed lower than the unification scale $M_{GUT}$.
\noindent At that point we would like to comment on different relations 
obtained in the literature. Witten's relation \cite{wi} reads
\begin{equation}
G_N = {({ \pi \over 4})}^{1/3} e^{-{2 \phi \over 3}} {{\alpha}_{GUT}^{4/3}
\over M_{GUT}^2}   
\end{equation}
and is obtained in our case by using ten-dimensional 
fivebrane units, which are
the natural units for $M$-theory, as we argued in the preceding
paragraph. However, the four-dimensional fivebrane units obtained through the 
Weyl rescaling (\ref{eq:ar23}) are probabaly more natural to use and
lead to (\ref{eq:ar26}). 
The analysis performed in \cite{da} uses a string metric. The
corresponding ten-dimensional Lagrangian is strongly coupled and cannot be, in
our opinion, used to derive four-dimensional relations.

Different approaches attempting to define four-dimensional
physical quantities have been discussed in detail in the literature and
can be found in \cite{dine}, \cite{ckm}.


\section{Conclusions}
The goal of this letter is to give a weak-coupling interpretation of
different terms in the strongly coupled $E_8 \times E_8$ heterotic
Lagangian by projecting onto ten-dimensions and keeping only zero modes.
This is possible by trading $M$-theory variables in terms of heterotic
string variables by using the membrane quantization condition. The
result is similar to a mixture of string tree-level and one-loop terms.
There are two distinct one-loop terms. The first is the usual
Green-Schwarz term, provided the anomaly cancellation mechanism at
strong coupling reduces to the usual one at weak coupling. This is
possible only with a non-minimal modification of the field 
strength of the three-form $C$ of the eleven-dimensional SUGRA 
( $\alpha=0, \beta=m=-1$ with the notations of Section $2$). The second
term has algebraically the required structure to be interpreted as part
of the heterotic one-loop $F^4$ corrections. 

In strong coupling, the Lagrangian is better expressed in dual,
fivebrane units. We argue that these units are more convenient for 
further compactification and for defining four-dimensional couplings and
scales.

There are certainly many issues which deserve further investigation.
One of them is to consistently integrate out the massive 
Kaluza-Klein modes, which give certainly additional contributions in the
Lagrangian describing the zero-modes. Another issue is the clarification of
the role played by the ten-dimensional antisymmetric tensor field in strong
coupling regime.

\vskip 1.2cm
{\bf Acknowledgements}
\vskip 0.3cm
We would like to thank S.P. de Alwis for useful comments and remarks.

\end{document}